# Implementation of Fuzzy Inference Engine for equilibrium and roll-angle tracking of riderless bicycle

Reza Yazdanpanah Abdolmalaki
Department of Mechanical, Aerospace and Biomedical Engineering
University of Tennessee, Knoxville
Email: ryazdanp@vols.utk.edu

*Abstract*—In this paper, a Fuzzy Inference System (FIS) is fabricated on a riderless bicycle. The Fuzzy Inference System is based on a rule base inherited from human experience of bicycle riding. The steady turning motion and roll-angle tracking controls for the riderless bicycle were achieved by using fuzzy concept. Collection of sensors, actuator, micro-controller and electrical circuits were employed to introduce new prototype autonomous bicycle. Effectiveness of the control scheme was proved by experimental tests and stabilization and roll-angle tracking of the real bicycle was illustrated by results.

## I. Introduction

It has been long time that the bicycle is established as a vehicle not only for transportation purposes, but also as a device for fun, sport, rehabilitation and lots of other aspects. Although lots of people have the experience of riding a bicycle and it is a well-known vehicle, its motion equation and the way of controlling a bicycle are still challenging problems. Bicycle has a complicated equation of motion which even now after long time there are innovative ideas about it.

Cognition of this vehicle, as the trend of developing in bicycle and motorcycle industry clearly showed, will result in introducing efficient vehicle. Bicycle is known as a useful tool for both environment and human health and lots of efforts were done in order to eliminate its weak points. As the main weakness of all the two-wheel vehicles which is their stability, control of bicycle has been a popular topic for researchers in the late half of the last century. During the early of 20th century, several authors studied the problems of self-stabilization, balancing, and steering.

Yamakita et al. utilized an input-output linearization method to design a trajectory tracking controller for an automatic bicycle [1], [2].

Several control structures have been proposed and implemented by Yazdanpanah et. al which can be used for modeling and controlling of bicycle [3], [4]. These methods also can be utilized in general control procedures such as in industrial, surgical and assistive robotics [5], [6].

Schwab et al. presented several approaches such as the pencil-and-paper, a numerical dynamics program and a symbolic software to drive the linear motion equation of bicycle. Its accuracy makes this model a benchmark [7].

Yavin also presented a nonlinear dynamic equation in which a simple structure of bicycle is used to develop equation of motion via the Lagrangian approach [8], [9], [10].

H. D. Sharna et al. introduced an intelligent controller for stabilizing an autonomous bicycle system [11]. The controller was developed by using fuzzy logic approach which the rule set was designed by using of the inherent-characteristic relationship of roll and steer present in a bicycle.

C. K. Chen and T.S Doa presented a steady turning and roll angle tracking control for unmanned bicycle with fuzzy controller by fixed parameters and rules [12]. Later they designed a Genetic fuzzy control for path tracking of an autonomous bicycle which optimized the membership function parameters with Genetic Algorithm (GA) [13].

Shafiekhani et al. designed and tested an adaptive neuro-fuzzy controller on an unmanned bicycle [14], [15] where the bicycle was controlled by steering angle.

In this paper, after giving a brief explanation about bicycle test setup and its equipments, the controller structure and its rule base are explained. Then results of the experimental tests are depicted which show that experimental results meet the expectation.

## II. Prototype equipped bicycle

In this study, Arduino Board has been used as a microcontroller which its duties are sending, receiving and operating data in order to balance the bicycle. Power supply has been considered to feed 5v and 24v for control drive circuit and DC motor, respectively. IMU sensor type MPU6050 measures the roll angle of bicycle ($\varphi$) and the encoder with 200 slots on its disc counts steering rotation ($\delta$). A DC motor with gearbox creates appropriate torque to turn the steering fork. Also, output shaft of gearbox has 130 rpm no-load speed at 24v with 2.5N.m stall torque. Moreover, position control of motor was done by PID controller which its coefficients were tuned by trial and error method ($k_p = 40$, $k_i = 0.001$, $k_d = 1$). It should be said that the utilized treadmill provides similar situation for running bicycle when it works on constant speed.

In order to eliminate noise from IMU data, 260 Hz accelerometer filtering and 256 Hz Gyro filtering have been set to sensor and Kalman filter also improved the accuracy of roll angle (figures 1, 2). Generally, the accelerometer is very noisy while the gyro drifts over time i.e., the gyroscope data can be used on a short term while the accelerometer data can be utilized on a long term. To treat this, a Kalman filter is then designed and implemented in which a combination of accelerometer and Gyro data are employed to achieve a better roll angle estimation along with removing unwanted residual noise.

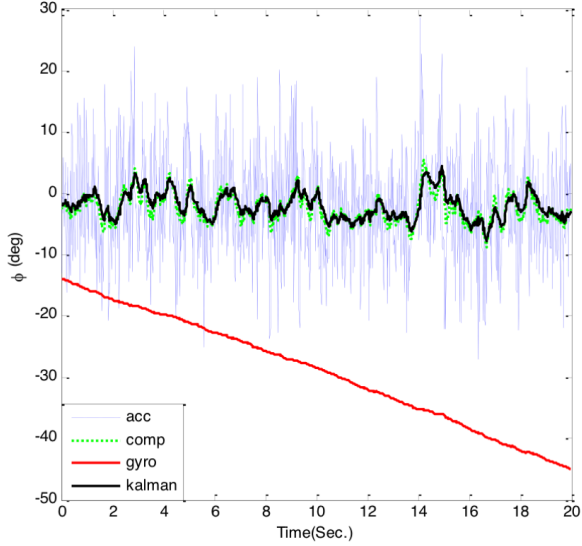

Figure 1. Roll-Angle of bicycle calculated by different filtering method

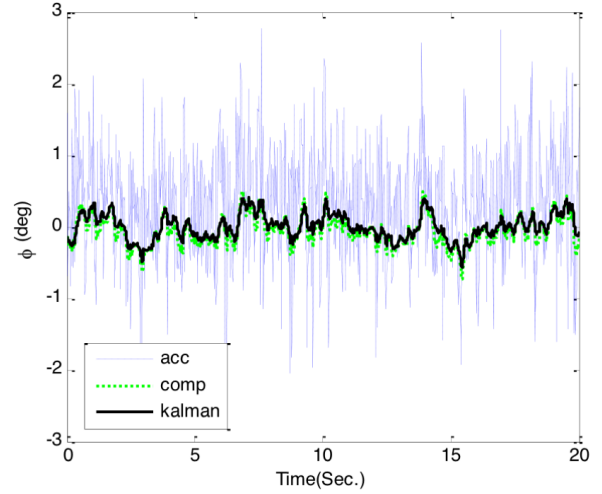

Figure 2. Roll-Angle of bicycle calculated by different filtering method after 260 Hz accelerometer and 256 Gyro filtering

The vibration isolator, which includes three different layers of foam, isolates the vibration of bicycle from IMU sensor. As it clearly is shown in figure 2, the Kalman filter has the lowest level of noise. Hence the Kalman filter was used for measuring the roll-angle of the bicycle.

The vibration isolator, which includes three different layers of foam, isolates the vibration of bicycle from IMU sensor. As it clearly is shown in figure 2, the Kalman filter has the lowest level of noise. Hence the Kalman filter was used for measuring the roll-angle of the bicycle.

## III. CONTROL STRUCTURE

In order to design a fuzzy controller, the rule bases which contain collection of if-then rules with properties that formulate human knowledge are needed. With considering these rules and choosing appropriate fuzzifier, defuzzifier and fuzzy inference engine, controller's structure was obtained and it could be implemented on a real bicycle.

### A. Balancing control

For balancing control of the unmanned bicycle on straight path, a control scheme was used which its structure is depicted in figure 3.

In this block diagram, Plant is the bicycle attitude, PID controller (inner loop) is a position controller which create the appropriate signal (Voltage) for DC motor placed on the fork. Outer control loop is the fuzzy inference system with roll angle and difference of it as inputs and reference steering angle as its output. Given the following situation, rule bases of fuzzy controller can be expressed.

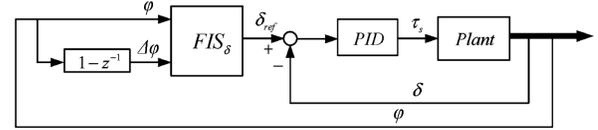

Figure 3. Balancing control block diagram

1) If $\varphi$ is negative large (NL) and $\Delta\varphi$ is negative large (NL), then $\delta$ is negative large (NL).
2) If $\varphi$ is zero (Z) and $\Delta\varphi$ is zero (Z), then $\delta$ is zero (Z).
3) If $\varphi$ is positive large (PL) and $\Delta\varphi$ is positive large (PL), then $\delta$ is positive large (PL).

where, positive direction of roll and steering angle is right side of bicycle.

Using the above approach, one could continue to formulate rules for all possible cases, as shown in table I.

For all fuzzy controllers in this study, triangular membership functions are used. These membership functions for inputs and output are illustrated in figure 4.

In this study, all membership functions were used from [12]

Table I
RULE-BASE FOR BALANCING FUZZY CONTROLLER

| $\varphi$ | NL | NM | NS | Z | PS | PM | PL |
|---|---|---|---|---|---|---|---|
| $\Delta\varphi$ | | | | | | | |
| NL | NL | NL | NL | NM | NM | NS | Z |
| NM | NL | NL | NM | NM | NS | Z | PS |
| NS | NL | NM | NM | NS | Z | PS | PM |
| Z | NM | NM | NS | Z | PS | PM | PM |
| PS | NM | NS | Z | PS | PM | PM | PL |
| PM | NS | Z | PS | PM | PM | PL | PL |
| PL | Z | PS | PM | PM | PL | PL | PL |

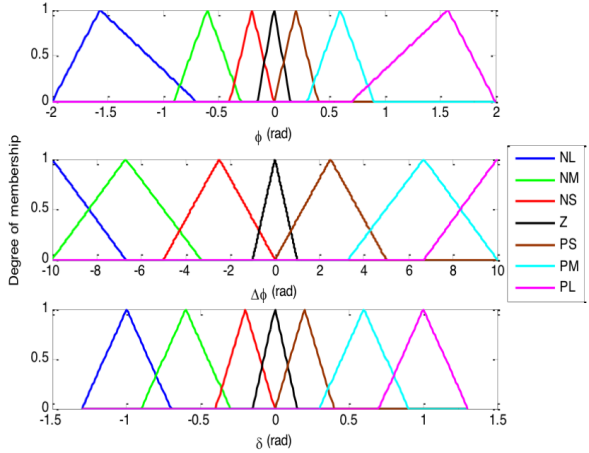

Figure 4. Membership functions for balancing FIS

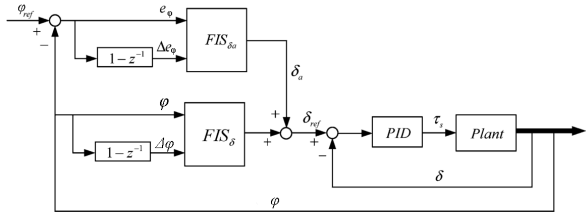

Figure 5. Equilibrium control block diagram

in which various numerical tests are performed by Chen et al. Mamdani combination has been used as a fuzzy inference engine with the following characteristics:

- And method: product
- Or method: probabilistic
- Implication: product
- Aggregation: sum
- Defuzzification: centroid

### B. Equilibrium and roll-angle tracking controls

Adding additional value δa to δref cause bicycle to balance with a roll angle (according to its velocity) and run on circular path. So, with establishing another fuzzy controller which gives an appropriate signal $\delta_a$, roll angle of unmanned bicycle can be controlled. figure 5 shows its block diagram.

In this block diagram, roll angle error $e_\varphi(k) = \varphi_{ref}(k) - \varphi(k)$ and rate of it $\Delta e_\varphi = e_\varphi(k) - e_\varphi(k-1)$ are inputs of the roll-angle tracking controller which gives $\delta_a$ on its output. Given the following situation, rule bases of fuzzy controller could be expressed.

1) If $e_\varphi$ is negative large (NL) and $\Delta e_\varphi$ is negative large (NL), then $\delta_a$ is positive large (PL)
2) If $e_\varphi$ is zero (Z) and $\Delta e_\varphi$ is zero (Z), then $\delta_a$ is zero (Z)
3) If $e_\varphi$ is positive large (PL) and $\Delta e_\varphi$ is positive large (PL), then $\delta_a$ is negative large (NL)

Table II
RULE-BASE FOR EQUILIBRIUM FUZZY CONTROLLER

| $e_\varphi$ | NL | NM | NS | Z | PS | PM | PL |
|---|---|---|---|---|---|---|---|
| $\Delta e_\varphi$ | | | | | | | |
| NL | PL | PL | PL | PM | PM | PS | Z |
| NM | PL | PL | PM | PM | PS | Z | NS |
| NS | PL | PM | PM | PS | Z | NS | NM |
| Z | PM | PM | PS | Z | NS | NM | NM |
| PS | PM | PS | Z | NS | NM | NM | NL |
| PM | PS | Z | NS | NM | NM | NL | NL |
| PL | Z | NS | NM | NM | NL | NL | NL |

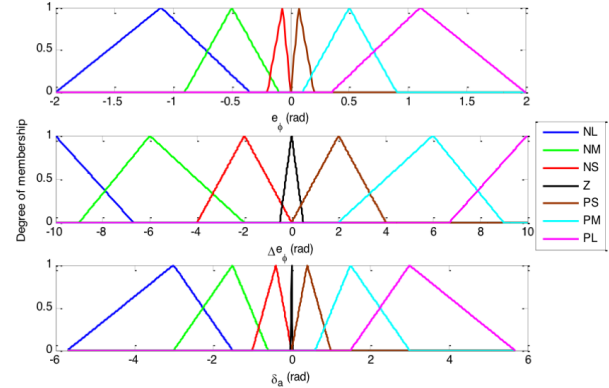

Figure 6. Membership functions for Equilibrium FIS

Using the above approach, one could continue to formulate rules for all possible cases, as shown in table II. These membership functions for inputs and output are illustrated in figure 6.

The fuzzy controller with the same specification for fuzzy inference engine as balancing controller has been used for roll-angle tracking control.

### IV. IMPLEMENTATION OF CONTROL ALGORITHM

The way of exchanging data between different equipments is shown in figure 7. From this figure, the measured roll angle by IMU sensor along with other data such as steering angle (measured by encoder) and voltage used by DC motor, are transferred to Arduino board. After some numerical computations like filtering and buffering of sensor data, the values of roll angle, encoder, time, voltage and PWM signal are sent to computer. This information is employed in MATLAB software and using them, suitable control signal is computed by controller. The controller output (reference steering angle) comes back to Arduino board and then using the motor driver (L298), the bicycle's fork is turned.

### V. EXPERIMENTAL TESTS

Two experimental tests (one for stability and another for roll-angle tracking) are performed to evaluate the effectiveness of the fabricated fuzzy controller. Figure 8 shows the bicycle response to initial condition $\varphi_0 = -20^0$ at constant velocity $v = 10 km/h$. One can see that the controlled roll angle is asymptotically decreased into ±5 degrees band.

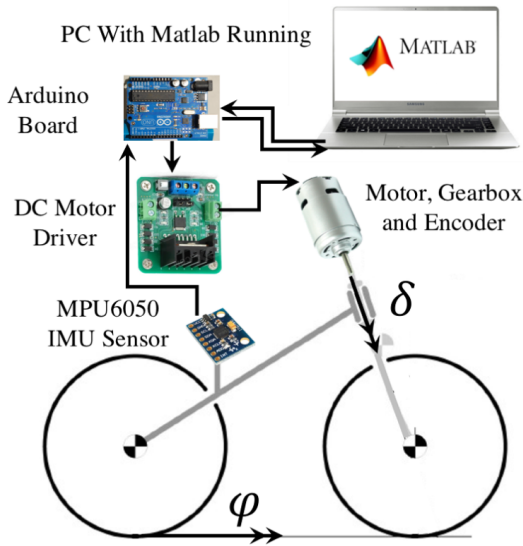

Figure 7. Exchanging data between different equipments

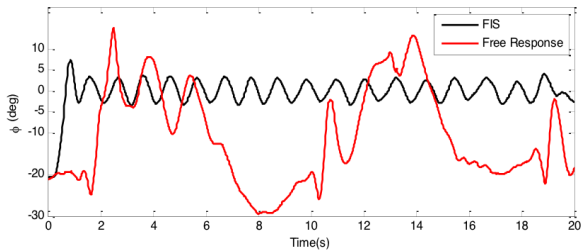

Figure 8. Bicycle response to initial condition

Comparing the control signal (as a set point) and actual steering angle (measured by encoder), as depicted in figure 9, shows a small difference in their values, which has reasons such as the error in measuring the roll-angle by IMU, motor inability to track reference angle and delays for sending, receiving and processing data. This difference causes the bicycle to oscillate five degrees around its perpendicular position.

In another test, by considering a sinusoidal function as the reference of the roll-angle tracking controller with amplitude of 50 and frequencies of π/3 rad/s, the response of the bicycle is illustrated in figure 10 in which the controller perfectly helped unmanned bicycle to track the reference roll-angle.

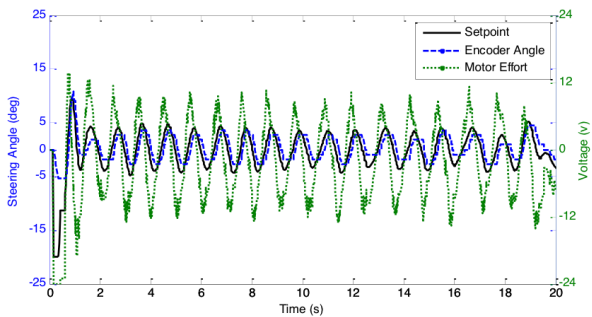

Figure 9. Controller and actuator response

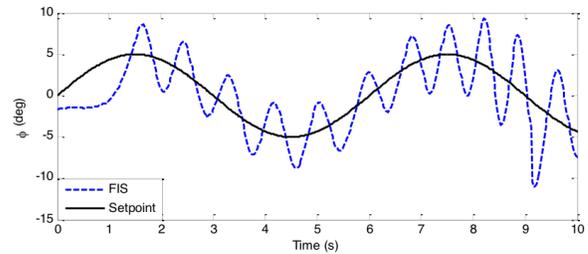

Figure 10. Bicycle response to reference sinusoidal roll-angle

## VI. CONCLUSION

In this study, a prototype equipped bicycle has been considered for implementing a fuzzy controller in order to control its balance and roll angle. According to this equipped bicycle, the fuzzy and PID controller have been designed to stabilize the bicycle in its straight-running motion and steering angle control, respectively. The experimental results indicate that the bicycle can be controlled to the upright position while face to initial condition. By considering another fuzzy controller for roll-angle tracking, bicycle could also follow the reference roll angle.